\documentclass[prd,aps,preprint,tightenlines,superscriptaddress]{revtex4}
\usepackage{mathrsfs}
\usepackage{amsfonts}
\usepackage{amsmath}
\usepackage{array}
\usepackage{verbatim}
\usepackage{epsfig}
\usepackage{graphicx}

\def    \o              {\ifmmode {\cal{O}} \else ${\cal{O}}$ \fi}
\def    \q              {\ifmmode {\cal{Q}} \else ${\cal{Q}}$ \fi}

\begin{document}
\title{Heavy Quarkonium Production at Low $P_\perp$  in NRQCD with Soft Gluon
Resummation}

\author{Peng Sun}
\affiliation{Center for High-Energy Physics, Peking University,
Beijing 100871, China}
\author{C.-P. Yuan}
\affiliation{Department of Physics and Astronomy, Michigan State University,
East Lansing, MI 48824, USA}
\affiliation{Center for High-Energy Physics, Peking University,
Beijing 100871, China}
\author{Feng Yuan}
\affiliation{Nuclear Science Division, Lawrence Berkeley National
Laboratory, Berkeley, CA 94720, USA}
\affiliation{Center for High-Energy Physics, Peking University,
Beijing 100871, China}
\


\begin{abstract}
We extend the non-relativistic QCD (NRQCD) prediction for the
production of heavy quarkonium with low transverse momentum in
hadronic collisions by taking into account effects from all order
soft gluon resummation. Following the Collins-Soper-Sterman
formalism, we resum the most singular terms in the partonic
subprocesses. The theoretical predictions of $J/\psi$ and
$\Upsilon$ productions are compared to the experimental data from
the fixed target experiments (E866) and the collider
experiments (RHIC, Tevatron, LHC). The associated non-perturbative
Sudakov form factor for the gluon distributions is found to be
different from the previous assumption of rescaling the quark form
factor by the ratio of color factors. This conclusion should be further
checked by future experiments on Higgs boson and/or di-photon
production in $pp$ collisions. We also comment on the
implication of our results on determining the color-octet matrix
elements associated with the $J/\psi$ and $\Upsilon$ productions
in the NRQCD factorization formalism.

\end{abstract}

 \maketitle

\section{introduction}
Heavy quarkonium production has been an important topic in strong
interaction physics where one  can apply perturbative QCD
calculations and systematically classify the associated non-perturbative physics
in heavy quark systems~\cite{Brambilla:2004wf}. An effective theory,
called non-relativistic QCD (NRQCD)~\cite{Bodwin:1994jh}, has been
developed to study the productions and decays of heavy quarkonia,
and especially their productions with large transverse momentum in hadronic collisions, up to the
next-to-leading order (NLO) in QCD interactions. These calculations
demonstrated that the so-called color-octet mechanism
is important to understand the $J/\psi$ and $\Upsilon$ productions
at the Fermilab Tevatron and the CERN Large Hadron Collider
(LHC)~\cite{Ma:2010yw,Butenschoen:2010rq,Chao:2012iv,Butenschoen:2011yh,Butenschoen:2012qh,Braaten:2000,Wang:2012is}.

However, how large of the associated color-octet matrix elements
remains a question. This becomes more important in light of recent studies
of the heavy quarkonium polarization in the collider experiments,
where two separate NLO calculations yield very different conclusions~\cite{Chao:2012iv}.
The main  difference lies on the size of the non-perturbative color-octet
matrix elements. In heavy quarkonium production,
the differential cross section is written into a factorized form,
\begin{eqnarray}
d\sigma=\sum_n d\hat{\sigma}(Q\bar{Q}[n]+X)\langle\o^H[n]\rangle \; .
\end{eqnarray}
Here $\hat{\sigma}(Q\bar{Q}[n]+X)$ is the cross section at the
parton level, which represents the production of a pair of heavy
quark in a fixed color, spin and orbital angular momentum state
$n$ and can be calculated perturbatively. The long distant matrix
element (LDME) $\langle\o^H[n]\rangle$ describes the transition of
the heavy quark pair in the configuration of $Q\bar{Q}[n]$ into
the final state heavy quarkonium. The LDMEs are
process-independent, which can be extracted from experimental data
or calculated from non-perturbative method, such as the  potential
model for predicting  heavy quarkonium states~\cite{Bodwin:2006dn}. In
addition, the LDMEs are organized in terms of the velocity $v$
expansion in the NRQCD framework. A fixed order perturbative
calculation is performed in orders of both the strong coupling
constant $\alpha_s$ and the power of the velocity for the
associated LDMEs. For example,  for $J/\psi$ production in $pp$
collisions, the differential cross sections depend on the
following three color-octet matrix elements:
\begin{equation}
\langle\mathcal {O}^{J/\psi}[{}^3S_1^8)]\rangle\; ,\langle\mathcal
{O}^{J/\psi}[{}^1S_0^8]\rangle\; ,\langle\mathcal
{O}^{J/\psi}[{}^3P_J^8]\rangle \ ,
\end{equation}
which are at the same order in the velocity expansion.
M.~Butenschoen and B.~A.~Kniehl used a fixed-order calculation to
extract the values of these three matrix elements by fitting to
several sets of transverse momentum ($P_\perp$) distributions of
$J/\psi$ produced in hadron-hadron collisions~\cite{Butenschoen:2011yh,Butenschoen:2012qh}.
In their fits, the experimental data points in small
and mediate $P_\perp$ region are all included.
They found that the extracted values of these three LDMEs strongly depend on the
selection of the low $P_\perp$ data sets, by imposing a lower limit on the $P_\perp$
of the data included in the fits,
cf. Tables II, III and IV of Ref.~\cite{Butenschoen:2012qh}.
On the other hand,
K. T. Chao et al demonstrated that $J/\psi$ production in high $P_\perp$ region
is only sensitive to two linear combinations of these three matrix
elements~\cite{Ma:2010yw}.
In brief, the above three LDMEs cannot be precisely determined from comparing
the current experimental data with a fixed-order perturbative calculation
in the framework of NRQCD.
Moreover, the values of these matrix elements
extracted in~\cite{Butenschoen:2011yh,Butenschoen:2012qh}
seem to be too large when compared to those extracted from the
inclusive cross section of  $J/\psi$ production  in $e^+e^-$
annihilation~\cite{Brambilla:2004wf,Butenschoen:2011yh,Butenschoen:2012qh}.

In this paper, we will extend the NRQCD prediction
for heavy quarkonium production to low transverse momentum region with a
resummation calculation.
This will provide additional information on the color-octet
matrix elements. Furthermore, comparing the differential cross section spectrum at low
$P_\perp$ and high $P_\perp$ could shed light on the underlying mechanisms
for heavy quarkonium production, and may also improve our understanding of the
heavy quarkonium polarization.
Some of the recent calculations for the production of  heavy quarkonia  with  very high $P_\perp$
could be found in Refs.~\cite{Kang:2011mg,Fleming:2012wy}.
The  low transverse momentum heavy quarkonium production has its own
interest in applying the perturbative QCD factorization formalism.  In the low
$P_\perp$ region ($\Lambda_{QCD}\ll P_\perp\ll M$), there are large
logarithms of the type $\alpha_s^m\ln^{2m-1}(M^2/P_\perp^2)$ in
high-order perturbative calculations.
Therefore, to obtain a reliable perturbative prediction, we have to resum
these large logarithms. In this work we follow the Collins-Soper-Sterman
(CSS) formalism~\cite{Collins:1984kg}. We will derive the relevant resummation
coefficients, by comparing the expansion of the resummation formula
with the fixed order perturbative calculations within the NRQCD formalism.
In particular, the most
singular contributions from the color-octet ${}^1S_0^{(8)}$ and ${}^3P_J^{(8)}$
channels will be resummed through the CSS formalism. Earlier work on
the soft gluon resummation for heavy quarkonium production in low
$P_\perp$ region has been performed in Ref.~\cite{Berger:2004cc} in a model similar to
the color-evaporation model for heavy quarkonium production. In this paper, we
will calculate the resummation in the NRQCD framework.

The original CSS resummation (also called the transverse momentum
resummation)  is derived for low transverse momentum
Drell-Yan lepton pair (vector Boson) production, and has been applied to
Higgs boson and di-photon productions in
hadron-hadron collisions and semi-inclusive hadron production
in deep-inelastic scattering (DIS) processes.
However, there have been few studies to extend the CSS resummation
to more complicated processes, such as the di-jet production in hadronic collisions.
Because the final state in these hard processes carries color, the soft gluon
radiation and resummation might be very different from that of Drell-Yan lepton
pair production in Ref.~\cite{Collins:1984kg}.
For example, using the threshold resummation formalism to describe the di-jet production
in hadron collisions, it was found that
the relevant Sudakov form factor has to be modified into
a complicated matrix form~\cite{Botts:1989kf,Kidonakis:1997gm}.
Although the kinematics of the transverse momentum resummation is different from that
of the threshold resummation,  similar matrix form may exist. (See, for example,
a recent calculation in the soft-collinear-effective-theory~\cite{Zhu:2012ts}.)
Interestingly, when we apply these analyses to the heavy quarkonium production in NRQCD,
we find that for heavy quarkonium production at low
transverse momentum, the matrix form will be simplified,
and can be written as a single exponential form factor
in the CSS resummation formalism.
The reason is as follows. There is only one color-configuration,
in either color-singlet or color-octet state, for heavy quarkonium
production in hadron-hadron collisions, and there is no mixing between them.
As a result, the soft gluon resummation for the color-singlet channel
will have the exact the same form as that presented in Ref.~\cite{Collins:1984kg}.
For the color-octet channel, the most important leading double logarithmic terms are
found to be the same as those for the color-singlet channel.
On the other hand, for the
sub-leading-logarithmic terms, there is an additional term
in the color-octet channel,
as to be explicitly shown in the one-loop
calculation of the differential cross section.
Following the same arguments made in Refs.~\cite{Botts:1989kf,Zhu:2012ts}
that QCD resummation calculation may be performed
for more complicated hard processes (e.g., dijet production) in hadron collisions,
we examine in this work the effect of QCD resummation to the
production of heavy quarkonium in hadron collisions.
It will be interesting to check the validity of the resummation formalism
employed here beyond the NLO, such as at two-loop order for heavy
quarkonium production in NRQCD framework.

An important ingredient of the resummation calculation, for
predicting low transverse momentum distribution of
heavy quarkonium produced via gluon-gluon fusion processes,
is the determination of the needed non-perturbative
Sudakov form factor.
This form factor was
previously assumed to be related to that of the quark fusion processes by
the ratio of color factors $C_A/C_F=9/4$~\cite{Balazs:2000wv}.
The latter was determined through a global analysis of
Drell-Yan (lepton) pair productions in hadron
collisions~\cite{Davies:1984sp,Ladinsky:1993zn,Landry:1999an,Berge:2004nt}.
Our results will show this assumption does not work for
heavy quarkonium production. Future experimental
data on top quark pair and Higgs boson productions shall provide additional information on
determining the non-perturbative Sudakov form factors for gluon-gluon fusion
processes.

We would like to emphasize that the transverse momentum resummation
for colored final state hard processes, including heavy quarkonium production
in the color-octet channel and heavy quark pair production in general,
are far less developed as compared to those for the production of
color neutral particles, such as Drell-Yan pair and Higgs boson.
A call for caution is needed when comparing
the form factors determined from this calculation
to those from Drell-Yan processes. Also, we have to
keep it in mind that there has been no general proof of factorization for
hardronic hard processes (such as dijet and heavy quark pair productions)
at low transverse momentum in hadron collisions.
We note, however, an interesting development on this issue~\cite{Mitov:2012gt},
which may provide a support for such factorization.

The paper is organized as follows. In Sec. II, we derive the
low $P_\perp$ behavior of heavy quarkonium production in NRQCD by
taking the limit of $P_\perp\ll M$. At one-loop order, the perturbative
corrections are shown to contain large logarithms.
We resum these large logarithms to all orders in Sec. III, following
the CSS formalism. The relevant coefficients are
obtained at the next-to-leading-logarithmic (NLL) level.
The numerical evaluations are carried out in Sec. IV, where we extract the
non-perturbative Sudakov form factor for the gluon-gluon fusion processes
and the values of the associated color-octet matrix elements in the NRQCD framework.
We conclude our paper in Sec. V.

\section{Low $P_\perp$ Behavior of Fixed Order Calculations}

By applying the NRQCD factorization formalism, heavy quarkonium
production in hadron-hadron collisions arises from the partonic processes,
\begin{equation}
a+b\to [Q\bar Q]+X\ ,
\end{equation}
where $a$ and $b$ stand for the partons from the incoming nucleons.
In high energy collisions, it is dominantly produced
via  the gluon-gluon fusion subprocess.
At leading order of $v$ and $\alpha_s$, $(Q\bar Q)$ pairs
are produced in the ${}^1S_0$ or ${}^3P_J$ configuration via
gluon-gluon fusion. They can be produced in either color-singlet or
color-octet states, {\it i.e.}, $gg\to Q\bar Q[{}^1S_0^{1,8}]$ or $Q\bar Q[{}^3P_J^{1,8}]$.
Finite $P_\perp$ arises from the real gluon emission processes.
As to be shown later, low $P_\perp$ heavy quarkonia are
dominantly produced via the ${}^1S_0^{1,8}$ and ${}^3P_J^{1,8}$ channels.
This is because in these channels the initial state gluon radiation contributes to
a singular power behavior $1/P_\perp^2$ in the low $P_\perp$ region, whereas all other channels
are power suppressed by $P_\perp/M$ in the limit of $P_\perp\ll M$.
For example, the $2\to 2$ subprocesses for $J/\psi$ and $\Upsilon$ productions
include the following channels,
\begin{eqnarray}
&&gg\to [Q\bar Q^{J/\psi,\Upsilon}]_1+g\ , \\
&&qg\to [Q\bar Q^{J/\psi,\Upsilon}]_2+q \ , \\
&&q\bar q\to [Q\bar Q^{J/\psi,\Upsilon}]_2+g \ ,
\end{eqnarray}
where $[Q\bar Q^{J/\psi,\Upsilon}]_1$ can be in ${}^1S_0^{8}$, ${}^3S_1^{1,8}$,
and ${}^3P_J^{1,8}$ configurations, while $[Q\bar Q^{J/\psi,\Upsilon}]_2$ can be in ${}^1S_0^{8}$, ${}^3S_1^{8}$,
and ${}^3P_J^{1,8}$. The differential cross sections for these channels have been previously calculated
in Refs.~\cite{Cho:1995vh,Petrelli:1997ge,Klasen:2003zn,Meijer:2007eb}.
The low $P_\perp$ behavior of these cross sections can be obtained
by proper expansion in powers of $P_\perp/M$, where $P_\perp$ and $M$
are transverse momentum and mass of the heavy quarkonium, respectively.
In this expansion, we only keep the leading order contribution,
and neglect all higher order terms of $P_\perp/M$. In the limit $P_\perp\rightarrow0$, only
some of these cross sections can reproduce
the lowest order terms that are proportional to $1/P_\perp^2$. This case appears when
 the respective $2\rightarrow1$ partonic subprocess, $ab\rightarrow[Q\bar Q^{J/\psi,\Upsilon}]$,
 exist.
Hence, in high energy hadron-hadron collisions,
the production cross section of heavy quarkonium can be approximated
as follows:
 \begin{eqnarray}
\sigma(J/\psi \, {\rm or} \, \Upsilon) &=&
\sigma\,(gg\rightarrow{}^1S_0^{8})
+
\sigma \, (gg\rightarrow{}^3P_0^{8})+
\sigma \, (gg\rightarrow{}^3P_1^{8})  \nonumber \\ & +&
\sigma \, (gg\rightarrow\chi_2({}^1P_2^{1})) \cdot {\rm Br}\, (\chi_2({}^1P_2^{1})\rightarrow J/\psi \,{\rm or}\, \Upsilon +\gamma)\,, \label{sigmabr}
\end{eqnarray}
where we have ignored the contribution from
$q\bar{q}$ initial state, $q\bar{q}\rightarrow{}^3S_1^{8}$, for
its relatively small parton density in high energy collisions.
Furthermore, the production channel $gg\rightarrow\chi_0({}^3P_0^{1})$
is not included in our calculation, for its small decay branch ratio (Br)
into $J/\Psi$ or $\Upsilon$.

In the limit of $P_\perp\ll M$, the differential cross section of the
gluon-gluon scattering process, $gg\rightarrow\mathcal {Q}^{[c]}+g$,
 can be expressed as
\begin{eqnarray}
\frac{d\sigma}{dyd^2P_\perp}|_{P_\perp\ll M}&=&\sigma_{0}(\q^{[c]})\frac{\alpha_sC_A}{2\pi^2}\int
f(x)dxf(x') dx'\frac{1}{P^2_\perp}\left[\frac{2(1-\xi_1+\xi_1^2)^2}{(1-\xi_1)_+}\delta(1-\xi_2)\right.\nonumber\\
&+&\frac{2(1-\xi_2+\xi_2^2)^2}{(1-\xi_2)_+}\delta(1-\xi_1)
+\left.\left(2\ln\frac{M^2}{P_\perp^2}-\delta_{8c}\right)
\delta(1-\xi_2)\delta(1-\xi_1)\right]\ , \label{ptxs}
\end{eqnarray}
where $y$ and $P_\perp$ are rapidity and transverse momentum
of heavy quarkonium, respectively.
$\xi_1=Me^{y}/x\sqrt{S}$, $\xi_2=Me^{-y}/x'\sqrt{S}$, and $\sqrt{S}$ is the
center-of-mass energy of the hadron-hadron collider.
$f(x)$ and $f(x')$ are parton distribution functions (PDFs).
$\mathcal {Q}^{[c]}$ represents ${}^1S_0^{1,8}$ or ${}^3P_J^{1,8}$ state,
and $\delta_{8c}=1$ or $0$ for color-octet or singlet channel production.
$\sigma_0$ is proportional to the leading order partonic cross sections, and
\begin{eqnarray}
\sigma_{0}(\q^{[{}^3P_2^{1}]})&=&\frac{64}{15}\frac{\alpha^2_s\pi^3}{M^7}
\langle\mathcal {O}[{}^3P_2^1)]\rangle \,,\nonumber\\
\sigma_{0}(\q^{[{}^1S_0^{8}]})&=&\frac{5}{12}\frac{\alpha^2_s\pi^3}{M^5}
\langle\mathcal {O}[{}^1S_0^8)]\rangle \,,\nonumber\\
\sigma_{0}(\q^{[{}^3P_0^{8}]})&=&5\frac{\alpha^2_s\pi^3}{M^7}
\langle\mathcal {O}[{}^3P_0^8)]\rangle \,,\nonumber\\
\sigma_{0}(\q^{[{}^3P_2^{8}]})&=&\frac{4}{3}\frac{\alpha^2_s\pi^3}{M^7}
\langle\mathcal {O}[{}^3P_2^8)]\rangle\ .
\label{treelevel}
\end{eqnarray}
It is interesting to compare the above expression to that for the
color-singlet scalar particle (such as
the Higgs boson) production. (See, for example, Ref.~\cite{Ji:2005nu}.)
We find that for the color-octet channel,
there is an additional term with $\delta(1-\xi_1)\delta(1-\xi_2)$, cf. Eq.~(\ref{ptxs}),
which generates sub-leading logarithmic contribution in the low $P_\perp$ region.
It is originated from the interference of initial and final (colored) state soft gluon
radiations, which is absent in the production via color-singlet channel.

In order to perform the resummation calculation at the next-to-leading logarithmic (NLL)
level, we have to Fourier transform the above expression
to the impact parameter space, and include the virtual diagram
contributions. The Fourier transformation into the impact parameter
$b$-space is defined as
$W(b)=\int d^2 P_\perp e^{-iP_\perp\cdot b_\perp}\frac{d\sigma}{dyd^2P_\perp}$.
In the impact parameter space, the logarithmic
term will yield a soft divergence in terms of $1/\epsilon^2$, where
$\epsilon=(4-D)/2$ is the dimensional regularization parameter in D-dimension space. This
soft divergence will be cancelled by the virtual diagram contribution.
After adding up the contributions from real emission and virtual diagrams,
we are left with collinear divergence in term of $1/\epsilon$.
In our calculation, we adopt  the modified minimal
subtraction ($\overline{\rm MS}$) scheme to regularize
the remaining collinear divergences.
Finally, the differential cross sections
for describing the production of heavy
quarkonia in the low $P_\perp$ region, via
$gg \rightarrow {}^1S_0^{8}, {}^3P_J^{1,8}$ channels,
can be expressed in the impact parameter space, within the NRQCD formalism,
as
\begin{eqnarray}
W^{(1)}(b,M^2)&=&\sigma_{0}(\q^{[c]})\frac{\alpha_sC_A}{\pi}\int dxdx' f(x)f(x')\left\{\left[\xi_1{\cal
P}_{gg}(\xi_1)\delta(1-\xi_2)\left(-\frac{1}{\epsilon}+\ln\frac{4e^{-2\gamma_E}}{\mu^2b^2}\right)
    \right.\right.\nonumber\\
    &&\left.+(\xi_1\rightarrow \xi_2)\right]+\delta(1-\xi_1)\delta(1-\xi_2)
\left[(b_0+\frac{1}{2}\delta_{c8})\ln\frac{b^2M^2}{4}e^{2\gamma_E}\right.\nonumber\\
&&\left.\left.-\frac{1}{2}\ln^2\left(\frac{M^2b^2}{4}e^{2\gamma_E}\right)
    -\frac{\pi^2}{6}+ \frac{B^{[c]}_{\q}}{C_A}\right]\right\}\ , \label{wb}
\end{eqnarray}
where $b_0=(\frac{11}{6}C_A-\frac{2}{3}T_Fn_f)/N_c$,
with $C_A=N_c=3$, $T_F=1/2$, and $n_f$ is the number of light quark flavors.
The expressions for
$B^{[c]}_{\q}$ can be found in Ref.~\cite{Petrelli:1997ge}, and
$f(x)$ represents gluon PDF. The gluon
splitting function is defined as ${\cal P}_{gg}(x)=
    \frac{x}{(1-x)_+}+\frac{1-x}{x}+x(1-x)
    +\delta(x-1)\frac{b_0}{2}$.

\section{All Order Resummation}

The presence of large double logarithmic corrections in Eq.~(\ref{wb})
is a generic feature in low $P_\perp$ differential cross section.
In order to have a reliable prediction for low $P_\perp$ heavy quarkonium
production, we perform an all order resummation in the CSS formalism.
For the color-singlet channel, there is no final state radiation because of
its colorless nature. High order soft gluon radiation can only come from
initial state, and the CSS resummation will follow the case of Drell-Yan
pair production.
For the color-octet channel, the final state carries color, and the soft
gluon radiation from the color-octet $[Q\overline Q]$ state will contribute
an additional soft factor to the Sudakov exponent
when factorizing the heavy quarkonium production
process in the low transverse momentum region.
This feature has been clearly demonstrated in the above fixed order
calculation, cf. the factor of $\delta_{c8}$ in Eqs.~(8,10).

Soft gluon resummation for hard processes with colored final state
in hadronic processes has been systematically studied in the literature, for example,
for the threshold resummation of heavy quark pair production in Ref.~\cite{Kidonakis:1997gm}.
Recently, a transverse momentum resummation has been carried out in the
soft-collinear-effective-theory approach~\cite{Zhu:2012ts}. In the following,
we will derive the similar formula for the heavy quarkonium production.
Since heavy quark pair production and heavy quarkonium production do share
similarity in the hard processes, we would like to pay close attention
to the calculations of Refs.~\cite{Kidonakis:1997gm,Zhu:2012ts} to check
if we can gain some insight for the resummation of soft gluon radiation in heavy
quarkonium production, which can be viewed as a special case of heavy quark
pair production. In NRQCD, the heavy quark pair is produced at short distance
with fixed color and spin configuration, in particular, the heavy quark and antiquark
share the same momentum (half of quarkonium momentum) in the non-relativistic limit.
From the calculations of Refs.~\cite{Kidonakis:1997gm,Zhu:2012ts}, for
heavy quark pair production in $pp$ collisions, it was found  that
additional soft gluon radiation
from the final state particles introduces a matrix form for the soft and hard factors.
The hard factor can be calculated from fixed order perturbative diagrams, whereas
the effect of soft gluon radiation can be resummed through the anomalous dimensions
associated with the soft factor~\cite{Kidonakis:1997gm,Zhu:2012ts}.
Both the hard and soft factors are expanded in the orthogonal color basis in the
hard processes. For example, for
$g_ag_b\to Q_i\overline Q_j$ process, the following color basis are
chosen~\cite{Kidonakis:1997gm}:
\begin{equation}
c_1=\delta^{ab}\delta_{ij}  \ , ~~    c_2=if^{ab c}T^c_{ij}    \ , ~~   c_3=d^{abc}T^c_{ij} \  ,
\end{equation}
where $T^c$ is the generator of $SU(3)$ in the fundamental representation,
$a,b$ and $i,j$ are initial state gluons and final state quark pair color indices,
respectively. Clearly, $c_1$ corresponds to the color-singlet final state, while
$c_2$ and $c_3$ to the color-octet final states.
The low $P_\perp$ production of heavy quarkonium in color-octet state
is dominated by the
${}^1S_0^{(8)}$ and ${}^3P_J^{(8)}$ channels, which only couple to
the $c_3$ color basis. In other words, if we follow the calculations
of ~\cite{Kidonakis:1997gm} and
decompose the hard factor into the above three color bases, we will
find that at the leading order, only $H_{33}$ is non-zero.
In additional, the one-loop virtual corrections from~\cite{Petrelli:1997ge} also show
that at the NLO, only $H_{33}$ is non-vanishing. We expect this to be true for higher
order calculations as well. The reason is that the production of the
color-octet ${}^1S_0$ and ${}^3P_J$ channels requires
the initial state color indices to be symmetric, and hence a non-vanishing
contribution can only come from the $c_3$ color basis.
More importantly, the anomalous dimension $\Gamma$,
which governs the soft factor contribution to the resummation
calculation, has been calculated
up to one-loop order in Ref.~\cite{Kidonakis:1997gm}.
Applying those results to the case of color-octet
heavy quarkonium production in the NRQCD framework,
we find that the anomalous dimension $\Gamma$
becomes diagonalized, and
\begin{equation}
\Gamma=\frac{\alpha_s}{\pi}\left(
           \begin{array}{ccc}
           0  & 0 & 0 \\
           0 & -\frac{C_A}{2} & 0 \\
           0 & 0 & -\frac{C_A}{2} \\
         \end{array}
       \right)  \ ,
\end{equation}
where we have only kept the real part of the matrix elements, and they
correspond to the partonic threshold limit with heavy quark pair
produced at rest, cf. Eq.(4.8) of  Ref.~\cite{Kidonakis:1997gm}.
We note that taking the partonic threshold limit with heavy quark
pair produced at rest resembles the dominant kinematics of
the heavy quarkonium production in NRQCD.
Similar results were also found in the
soft-collinear-effective-theory calculations~\cite{Zhu:2012ts}.
After solving the renormalization group equation, we will obtain
an additional soft factor in the CSS resummation formula, as compared to that for
colorless particle, such as Higgs boson, production.
This conclusion also agrees with our explicit calculations.

The above analysis indicates that there is no complicated structure in the resummation
calculation for heavy quarkonium production in either color-singlet or color-octet channel.
Therefore, we can directly follow the CSS method to derive its final result.
In particular, from Eq.~(\ref{wb}) we can write down a differential equation with
respect to $\ln M^2$:
\begin{equation}
\frac{\partial W(b,M^2)}{\partial \ln M^2}=(K+G') W(b,M^2) \ ,
\end{equation}
where $K$ and $G'$ are soft and hard evolution kernels, and at one-loop order,
we have
\begin{equation}
K+G'=-\frac{\alpha_sC_A}{\pi}\ln
\left(\frac{M^2b^2}{4}e^{2\gamma_E-b_0-\frac{\delta_{c,8}}{2}}\right) \ .
\end{equation}
 The soft part $K$ depends on the scale $1/b^2$ and the renormalization scale
$\mu$, while $G'$ depends on the hard scale $M^2$ and $\mu$. Compared to that in
Higgs boson production, the only difference is the
additional term $\frac{\delta_{c,8}}{2}$, which arises
from the interference of the initial and final state soft gluon radiation
in the color-octet channel and is absent in the color-singlet channel.
Both the soft and hard parts $K$ and $G'$ obey the
renormalization group equation~\cite{Collins:1984kg},
\begin{equation}
\frac{\partial K}{\partial \ln\mu}=-\frac{\partial G'}{\partial \ln
\mu}=-\gamma_{Kg}=-2\frac{\alpha_sC_A}{\pi} \ ,
\end{equation}
where $\gamma_{Kg}$ is the cusp anomalous dimension.
After solving the renormalization group equations, and evolving from the low energy scale
of ${\cal O}(1/b)$  to  the high energy scale of ${\cal O}(M)$, we obtain,
\begin{eqnarray}
W(b,M^2)=e^{-{\cal S}_{Sud}(M^2,b,C_1,C_2)}
W(b,C_1,C_2) \ ,
\label{wpiece}
\end{eqnarray}
where the Sudakov form factor is
\begin{equation}
{\cal S}_{Sud}=\int_{C_1^2/b^2}^{C_2^2M^2}\frac{d
\mu^2}{\mu^2}\left[\ln\left(\frac{C_2^2M^2}{\mu^2}\right)
A(C_1,\mu)+B(C_1,C_2,\mu) \right]\ .
\end{equation}
Here $C_1$ and $C_2$ are two free parameters, at the order of unity. The functions
$A$ and $B$ can be expanded perturbatively in powers of $\alpha_s$, with
$A=\sum\limits_{i=1}^\infty
A^{(i)}\left(\frac{\alpha_s}{\pi}\right)^i$ and
$B=\sum\limits_{i=1}^\infty
B^{(i)}\left(\frac{\alpha_s}{\pi}\right)^i$.
Furthermore, in Eq.~(\ref{wpiece}),
\begin{eqnarray}
W(b,C_1,C_2)&=&\sigma_{0}(\q^{[c]})\frac{M^2}{S}\int\frac{dx}{x}\frac{dx'}{x'}
C_{gg}\left(\frac{x_1}{x},b,C_1,C_2,\mu^2\right)
C_{gg}\left(\frac{x_2}{x'},b,C_1,C_2,\mu^2\right)\nonumber\\
&& \times f(x,\mu) f(x',\mu) \ ,\label{cgluon}
\end{eqnarray}
where $x_1=Me^y/\sqrt{S}$, $x_2=Me^{-y}/\sqrt{S}$, and the scale $\mu=C_3/b$.
After choosing $C_1=C_3=2e^{-\gamma_E}$ and $C_2=1$, we obtain, for the channels
$gg \rightarrow {}^1S_0^{[c]}, {}^3P_J^{[c]}$:
\begin{eqnarray}
A^{(1)}&=&C_A,~~~B^{(1)}=-(b_0+\frac{1}{2}\delta_{c8})C_A \ , \nonumber\\
C_{gg}^{(1)}&=& \delta(1-x) \ , \nonumber\\
C_{gg}^{(1)}&=&\left(-\frac{\pi^2}{12}C_A+
\frac{B^{[c]}_{\q}}{2}\right)\delta(1-x), ~~~C_{gq}^{(1)}=\frac{C_F}{2}x\ .
\end{eqnarray}
Together with
$A^{(2)}= C_A \left[ \left( \frac{67}{36} -\frac{\pi^2}{12}
\right)N_c - \frac{5}{18}N_f \right] $, which is the same as that in the $gg\rightarrow H$ process~\cite{Cao:2009md},
the above result contains all the needed coefficients for performing the
 resummation calculation  at the next-to-leading logarithmic level for the
heavy quarkonium production in hadron collisions.
We note that both  $B^{(1)}$ and
$C^{(1)}$ for color-octet channels are different from those for color-singlet channels.
The additional contribution in $B^{(1)}$ for the color-octet channel is
exactly the same as that found in Refs.~\cite{Kidonakis:1997gm,Zhu:2012ts}
for studying the production of heavy quark pair in the threshold resummation
formalism.

It is interesting to note that the numerical values of the $C_{gg}^{(1)}$ coefficients for the color-octet
channels are not very different from each other. With the given $B^{[c]}_{\q}$ values from
Ref.~\cite{Petrelli:1997ge}, we obtain
\begin{eqnarray}
C_{gg}^{(1)}({}^1S_0^{8})=3.16,~~
C_{gg}^{(1)}({}^3P_0^{8})=3.76,~~
C_{gg}^{(1)}({}^3P_2^{8})= 2.80 \ .
\end{eqnarray}
Consequently, the ratio between the color-octet ${}^3P_J$ and
${}^1S_0$ contributions, derived from  Eq.(\ref{treelevel}), can be approximated as
\begin{eqnarray}
&&\left(C_{gg}^{(1)}({}^3P_0^{8})\sigma_{0}(\q^{[{}^3P_0^{8}]})+C_{gg}^{(1)}({}^3P_2^{8})\sigma_{0}(\q^{[{}^3P_2^{8}]})\right):
\left(C_{gg}^{(1)}({}^1S_0^{8})\sigma_{0}(\q^{[{}^1S_0^{8}]})\right) \nonumber\\
 =&&\frac{7.1}{M_Q^2}\frac{\langle\o[{}^3P_0^8]\rangle}{\langle\mathcal
 {O}[{}^1S_0^8)]\rangle}\approx \frac{7}{M_Q^2}\frac{\langle\o[{}^3P_0^8]\rangle}{\langle\mathcal
 {O}[{}^1S_0^8)]\rangle} \ ,
\end{eqnarray}
where $M_Q$ is the heavy quark mass, and $M_Q=M/2$ in NRQCD factorization formalism.
Here, we have simplified the two $P$-wave color-octet contributions by applying the
heavy quark symmetry $\langle\o[{}^3P_2^8]\rangle=5\langle\o[{}^3P_0^8]\rangle$.
Therefore, at the NLO accuracy, the total
contribution from the above three channels are nearly proportional
to the linear combination of color octet matrix elements:
\begin{eqnarray}
\sigma(\q^{[8]})\sim
\left(\langle\o[{}^1S_0^8]\rangle+\frac{7}{M_Q^2}\langle\o[{}^3P_0^8]\rangle
\right)\ , \label{combination}
\end{eqnarray}
which is the same as that goes into the leading order cross section calculation.
The above combination will enter into the cross section calculation of the low $P_\perp$ heavy quarkonium
production. This combination is different from that needed for calculating the high $P_\perp$
distributions~\cite{Ma:2010yw,Butenschoen:2010rq,Wang:2012is}.
This difference mainly comes from the fact that the effective
gluon-gluon-heavy-quark-pair coupling varies with kinematics.
At low $P_\perp$, we can take heavy quark mass limit
($M\gg P_\perp$), where both gluons are almost on-shell. At large $P_\perp$,
one of the gluon in the $2\to 2$ subprocesses must be far off-shell (at order of $P_\perp$),
and the gluon-gluon-heavy-quark-pair coupling will be different from that
in the low $P_\perp$ region.
These changes depend on the configuration of the heavy quark pair
and higher order corrections.
However, in the low $P_\perp$ region, the
soft (and collinear) gluon radiation dominates, and the differential
cross section yields a result proportional to the Born level contribution, cf. Eq.(\ref{combination}).
This has been confirmed by the exact NLO calculation which includes all the
$2\to 3$ subprocesses~\cite{Ma:2010yw,Butenschoen:2010rq}.

\section{numerical calculation}

In the numerical calculation, the differential cross section can be written as:
\begin{equation}
d\sigma=d\sigma^{resum}+(d\sigma^{pert}-d\sigma^{asym})\ ,
\end{equation}
where $\sigma^{resum}$ is the resummation of terms which are
proportional to $1/P_\perp^2$ in each order of perturbative
calculation, and $\sigma^{asym}$ is constructed to cancel the
same terms in the fixed-order cross section $\sigma^{pert}$. Therefore,
the second term $(d\sigma^{pert}-d\sigma^{asym})$ is power suppressed by
$P_\perp/M$ in low $P_\perp$ region. The above expression is in principle
valid in the whole transverse momentum region. In this paper,
we focus on the low $P_\perp$ region, where we can safely neglect
the contribution from the second term.

In the CSS resummation formalism, the resummation part of total cross section can be written as
\begin{equation}
\label{WY} {\frac{d\sigma }{ d^{2}P_{\perp}dy\, \, }}|_{P_\perp\ll M}=
{\frac{1}{(2\pi )^{2}}} \int d^{2}b\, e^{i{\vec{P}_{\perp}}\cdot
{\vec{b}}}{W(b,M,x_{1},x_{2})}\ ,
\end{equation}
where $W(b,M,x_{1},x_{2})$ has been extensively discussed in the last section.
The $b$ integral contains contribution from the
non-perturbative region where $b$ is so large that $\alpha_s(1/b)$
cannot be reliably calculated perturbatively.
To model the contribution from the non-perturbative region,
we follow the $b_*$-prescription to
add a phenomenological non-perturbative form factor~\cite{Collins:1984kg},
and write
\begin{equation}
\label{wpres} W(b)=W(b_{*}) W^{NP}(b)\, ,
\end{equation}
where $b_*$ is defined as $b_{*}={{b}/{\sqrt{1+(b/b_{max})^{2}}}}$.
Here, $b_*$ cannot exceed $b_{max}$, which is equivalent to
making a cutoff on the variable $b$ at $b_{max}$.

Generally, the non-perturbative form factor $W^{NP}$
depends on the flavor of the initial state partons. There have been
many studies in the literature to extract $W^{NP}$ associated
with the initial state quarks by comparing theoretical predictions with
experimental data of Drell-Yan lepton pair production
and $Z^0$ and $W$ boson productions in $pp$ ($p\bar p$)
collisions.
(See, for example, Refs.~\cite{Davies:1984sp,Ladinsky:1993zn,Landry:1999an}.)
On the other hand, for the gluon initiated processes, there has been no precise
determination of the corresponding $W^{NP}(b)$ in the literature.
For example, in Ref.~\cite{Balazs:2000wv}, di-photon productions via
gluon-gluon fusion processes in hadron collisions have
been investigated, where
the associated $W^{NP}(b)$ factor was assumed to scale with the color-factor
$C_A/C_F=9/4$, with respect
to that for the quark initiated processes~\cite{Berge:2004nt}.
This is because at the Tevatron energy, the production rate of
di-photon events is dominated by quark initiated subprocesses,
so that it is difficult to use that data to extract the
non-perturbative factor $W^{NP}(b)$ associated with the gluon
initiated subprocesses.
 In this work,
we will take the BLNY parameterization form, a 3-parameter pure Gaussian
form, as  proposed  in Refs.~\cite{Ladinsky:1993zn,Landry:1999an,Berge:2004nt},
\begin{equation}
\label{BLNY_form}W^{NP}(b)= {\textrm{exp}}\left[ -g_{1}-g_{2}\ln
\left( {\frac{Q}{2Q_{0}}}\right) - g_{1}g_{3}\ln
{(100x_{1}x_{2})}\right] b^{2}\ ,
\end{equation}
and extract the non-perturbative
form factor associated with the gluon initiated processes
by performing a global fit to the low $P_\perp$ distributions of
heavy quarkonia produced in high energy hadron-hadron
collisions.
More specifically, we will follow the previous studies and take the following
fixed parameters: $ Q_{0}=1.6\mbox {\, GeV} $ and $ b_{max}=0.5\mbox {\,
GeV}^{-1} $, in addition to the three free parameters $g_1$,
$g_2$, and $g_3$.
Furthermore, for simplicity, we will assume the same non-perturbative function for calculating both the
color-octet and color-single heavy quarkonium state productions.
This approximation is justified by that the low
 $P_\perp$ heavy quarkonia are dominantly produced  at high energy colliders via color-octet channel.

\begin{figure}[t]
\centering
\includegraphics[width=1\textwidth]{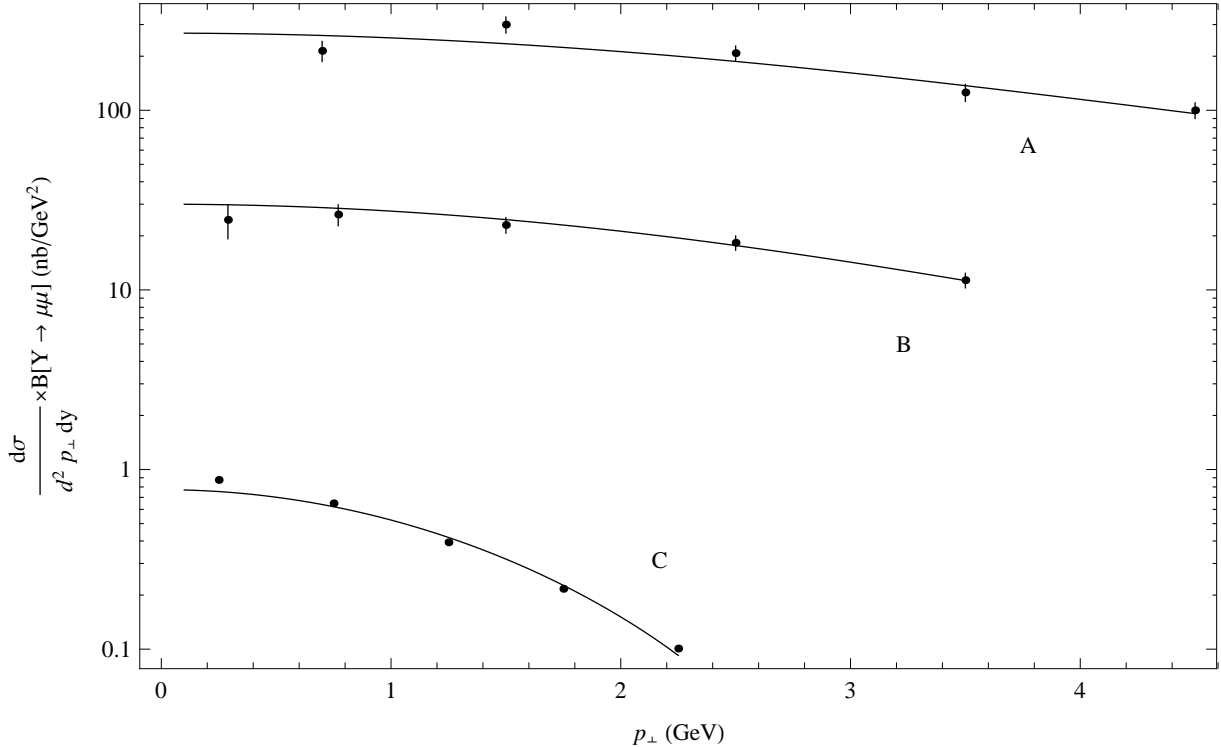}%
 \caption{\small $\Upsilon$
production at $\sqrt{S}=$7 TeV and $|y|<2$ at the LHC (A), at
$\sqrt{S}=$1.8 TeV and $|y|<0.4$ at the Tevatron (B), and at
$\sqrt{S}=$38.8 GeV and $-1<x_F<1$ for $pp$ collision at E866 (C).
The data points are from Refs.~\cite{:2007mja,Khachatryan:2010zg,Acosta:2001gv} for E866,
CMS, and CDF Collaborations, respectively.} \label{nUABC} \vspace{-0mm}
\end{figure}
\begin{figure}[t]
\centering
\includegraphics[width=1\textwidth]{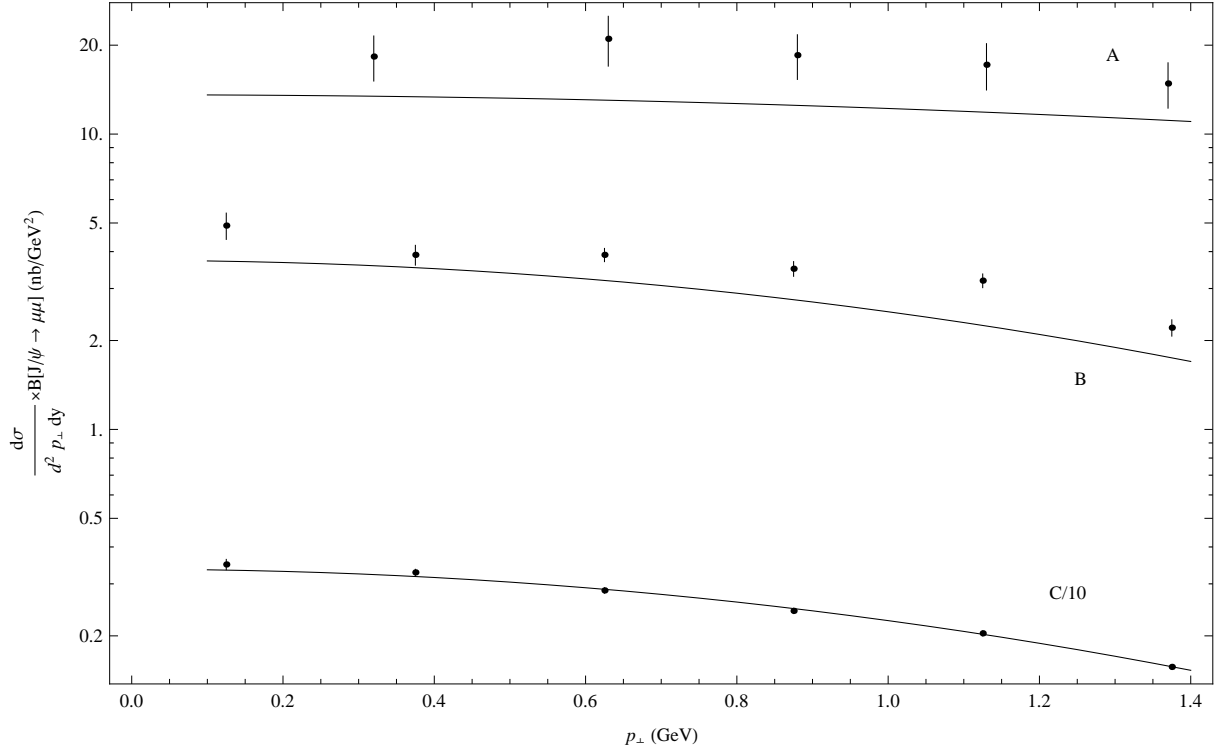}%
\caption{\small $J/\psi$ production at
$\sqrt{S}=$ 7 TeV with $1.6<|y|<2.4$ at the LHC (A), at $\sqrt{S}=$200 GeV
with $|y|<0.35$ (B) and $1.2<|y|<2.4$ (C ) for $pp$ collision at the RHIC.
The curve C has been multiplied by a factor $1/10$ to separate it from the others.
The data points are from Refs.~\cite{Khachatryan:2010yr,Adare:2011vq} for CMS
and PHENIX Collaborations, respectively.}
\label{nJABC} \vspace{-0mm}
\end{figure}
\begin{figure}[t]
\centering
\includegraphics[width=1\textwidth]{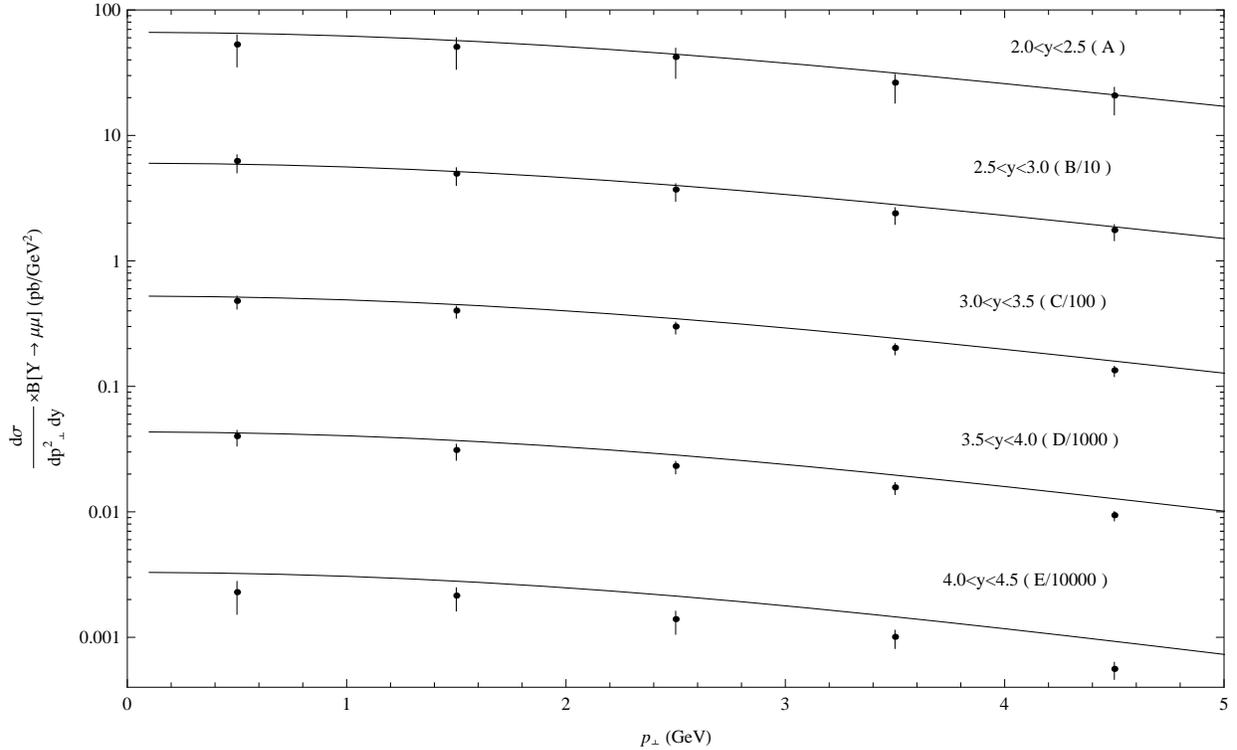}%
 \caption{\small $\Upsilon$
production at $\sqrt{S}=$7 TeV from the LHCb Collaboration.
The cures A, B, C, D, and E correspond to five rapidity bins: $2<y<2.5$,
$2.5<y<3$, $3<y<3.5$, $3.5<y<4$, and $4<y<4.5$.
To separate these cures in one figure, we
multiply curves B, C ,D, and E  by factors $1/10$, $1/100$, $1/1000$,
and $1/10000$, respectively. The data points are from Ref.~\cite{:2012ve}
for the LHCb Collaboration.} \label{uLHCb} \vspace{-0mm}
\end{figure}
\begin{figure}[t]
\centering
\includegraphics[width=0.7\textwidth]{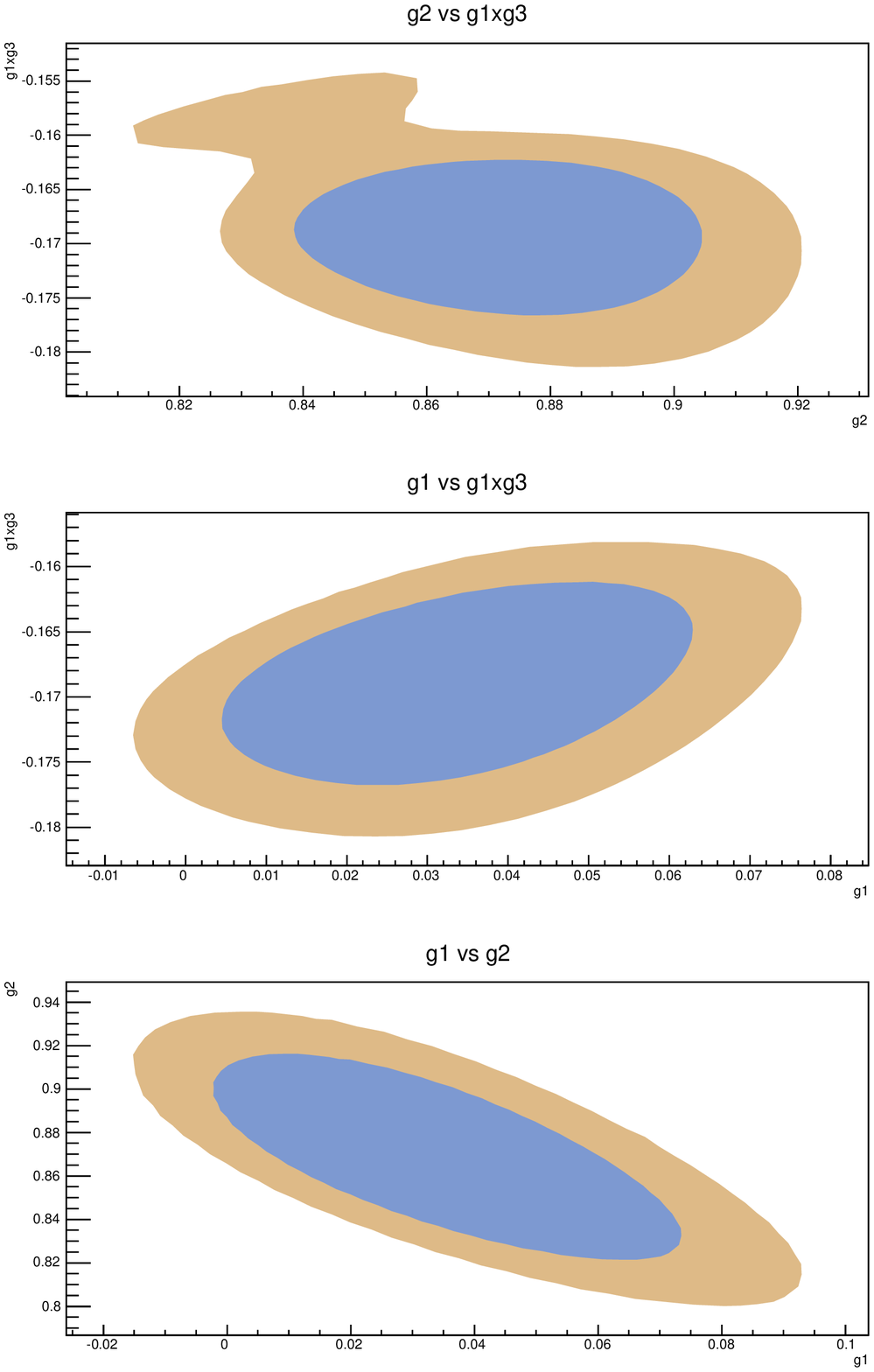}%
 \caption{\small The uncertainty contour for $g_1$, $g_1$ and $g_1\times g_3$. The yellow and blue region
 correspond to 90\% CL and 68\% CL, respectively
  } \label{contour1} \vspace{-0mm}
  \label{contour}
\end{figure}

As we discussed in the last section, heavy quarkonium (charmonium
and bottomonium) production rate depends on the associated color-octet
matrix elements in the NRQCD factorization formalism. In particular, we found that
the differential cross section in the low $P_\perp$ region is proportional to the combination
of the two hadronic matrix elements, as shown in Eq. (\ref{combination}). Hence, we shall treat the very
combination of $\q^{[{}^1S_0^{8}]}$ and $\q^{[{}^3P_0^{8}]}$ as one free parameter to be determined
 by the global fits to the $J/\psi$ and $\Upsilon$ production data, separately. We note that the
 hadronic matrix elements are different for $J/\psi$ and $\Upsilon$ productions.
  In our fits, we include the experimental data on $\Upsilon$ production from fixed
target experiment by the E866 Collaboration~\cite{:2007mja}, and the
collider experiments on $\Upsilon$ production at the
LHC~\cite{Khachatryan:2010zg,Khachatryan:2010yr,:2012ve} and the
Tevatron~\cite{Acosta:2001gv}, and $J/\psi$ production at the RHIC~\cite{Adare:2011vq} with $1.2<|y|<2.4$. The
other color-singlet matrix elements,  $\langle\mathcal {O}^{
J/\psi}[{}^3P_2^1)]\rangle$ and $\langle\mathcal {O}^{\Upsilon
}[{}^3P_2^1)]\rangle$, are taken from Ref.~\cite{Ma:2010yw} and
Ref.~\cite{Braaten:2000} for $J/\psi$ and $\Upsilon$ production, respectively.
In total, our fits contain ten free parameters: $g_{1,2,3}$, two linear
combinations of color-octet matrix elements for $J/\psi$ and
$\Upsilon$ production, respectively, and five normalization parameters $N_{fit}$ which were introduced
 to account for the normalization
uncertainty in each experiment, following the same procedure as
done in Ref.~\cite{Landry:1999an}.

To determine the free parameters $g_{1,2,3}$ and the two linear combination of color-octet matrix
elements in the global $\chi^2$ fits, we minimize the $\chi^2$ contribution from the five experimental
data sets, as shown in Table I, while allowing the normalization of each experimental data set to float within one
 standard deviation of the published experimental error of collider luminosity.
 The result of our fits are shown in Figs.~1, 2 and 3, in which we have plotted the shifted
 theory prediction (multiplied by $N_{fit}$ for Figs.~1 and 2) with the experimental data for the above five experimental data sets.
 In addition, we also compare the theory prediction, with the fitted theory parameters, to the $J/\psi$ production
 at the RHIC with $|y|<0.35$ (cf. Fig. 2(B)) and to the $\Upsilon$ production at the LHCb (cf. Fig. 3). The fitted parameters are
 summarized in Table I. The uncertainties quoted for
$g_{1,2,3}$ are evaluated with the other parameters fixed at their values
given by the best fit.  The overall
agreement between the theoretical predictions, based on the NRQCD and
soft gluon resummation, and the experimental data is very good,
considering the energy span of the experiments, ranging from fixed target to
high energy collider experiments. In particular, if we compare the experimental data from
the Tevatron and the LHC with those from the fixed target experiment, cf. Fig.~.1, we
find that there is a strong energy dependence for the
$P_\perp$-spectrum, which is reasonably described by the energy
dependence of the non-perturbative form factor in
Eq.~(\ref{BLNY_form}). Similar conclusion holds for the $J/\psi$
production between RHIC and LHC, as shown in  Fig.~2.
From Figs.~2 and 3, we find that the agreement
between theoretical predictions  and experimental data  for $J/\psi$ production is not as good
as that for $\Upsilon$ production. This may be because the mass of
$J/\psi$ is not very large (as compared to $\Lambda_{QCD}$) and the theoretical uncertainties
from the non-perturbative factor of Eq.~(\ref{BLNY_form})
become sizable at relative low mass region.

\begin{table}[t]
\caption{The results of the fits at 68\%, 90\% and 95\% CL. Here,
\protect$ N_{fit}\protect$ is the fitted normalization factor for each
experiment, which is multiplied to the corresponding theoretical prediction
to yield the comparison presented in Figs.~1 and 2.
  } {\centering
\begin{tabular}{|c|c|} \hline Parameter &
 BLNY fit \\
\hline \hline $ g_{1} $ (\textmd{CL}\; 68\%)&
0.03$\pm0.056$ \\
\hline $ g_{2} $ (\textmd{CL}\; 68\%) &
 0.87$\pm0.065$ \\
\hline $ g_{3}*g_1$  (\textmd{CL}\; 68\%)&
-0.17$\pm0.011$ \\
\hline \hline $ g_{1} $ (\textmd{CL}\; 90\%) &
0.03$\pm 0.111 $ \\
\hline $ g_{2} $  (\textmd{CL}\; 90\%) &
 0.87$\pm0.080$ \\
\hline $ g_{3}*g_1$ (\textmd{CL}\; 90\%)  &
-0.17$\pm0.013$ \\
\hline \hline $ g_{1} $ (\textmd{CL}\; 95\%)  &
0.03$ \pm0.113$ \\
\hline $ g_{2} $ (\textmd{CL}\; 95\%) &
 0.87$\pm0.134$ \\
\hline $ g_{3}*g_1$  (\textmd{CL}\; 95\%) &
-0.17$\pm0.030$ \\
\hline \hline E866 ($\Upsilon$) &
 $ N_{fit}=1.06$  \\ (5 points)
 & $\chi^2=7.9$
\\
\hline Tevatron ($\Upsilon$) & $ N_{fit}= 0.96$ \\ (5 points)
  & $\chi^2=3.0$
\\
\hline CMS($\Upsilon$) &
  $ N_{fit}=1.05 $\\ (5 points)
 & $\chi^2=9.2$
\\
\hline RHIC($J/\psi$) &
 $ N_{fit}=0.9$ \\ (6 points) ($1.2<|y|<2.4$)
  & $\chi^2=3.4$\\
 \hline CMS($J/\psi$ ) &
  $ N_{fit}=1.11$\\ (5 points)
 & $\chi^2= 14.6$
\\
\hline \hline $ \chi ^{2} $&
 38  \\
 \hline \hline $ \chi ^{2} /$DOF &
 1.46  \\
\hline
\end{tabular}\par}
{\centering \label{fit_result}\par}
\end{table}

The above resummation results are
based on the NRQCD factorization formalism for heavy quarkonium production.
Hence, the resulting non-perturbative
factor for the gluon-gluon initiated low
$P_\perp$ heavy quarkonium production
processes , cf. Eq.~(\ref{BLNY_form}), may not be the same as that for the
gluon-gluon initiated Higgs boson production in hadron-hadron collisions.
The former is mainly produced via color-octet channel, based on the
NRQCD factorization formalism, while the latter is produced via color-singlet
channel.
Although the resummed cross section has properly taken into account
the different effects of multiple soft gluon
radiation for producing a color-octet or singlet state in perturbative calculation,
a different non-perturbative factor might be needed to describe the Higgs boson
production, as compare to the heavy quarkonium production via color-octet channel.
 Nevertheless, we shall compare our
result to the assumption made in Ref.~\cite{Landry:1999an} for Higgs boson production,
\begin{eqnarray}
g_1=0.21\times (9/4)=0.47, ~
   g_2=0.68\times (9/4)=1.53,~
   g_1\times g_3=-0.29 \ ,
   \label{gblny}
\end{eqnarray}
which were scaled by the ratio of color factors involved in the gluon-gluon versus quark-quark fusion processes.
From our fit, we find that the absolute
values for the two parameters: $g_2$ which controls the scale dependence ($\ln Q$ term),
and the combination $g_1\times g_3$ which controls
the energy dependence ($x_1x_2$ term), are both smaller
than those in Eq.~\ref{gblny}
It will be interesting to further investigate this
important issue and the relevant phenomenological consequences.
In particular, our determinations of the above parameters are only based on heavy
quarkonium production in $pp$ collisions, where the NRQCD factorization contains
color-octet channel contributions and the associated non-perturbative form factor
might be different from that for the color-singlet processes.
To further check the scaling proposed in Ref.~\cite{Landry:1999an}, we need to
study the transverse momentum distribution of Higgs boson and/or di-photon
production in $pp$ collisions.
Finally, in order to estimate the uncertainties in our fit, we show in Fig.~\ref{contour}
the contour plots of the uncertainties of $g_1$, $g_2$, and $g_1\times g_3$.

Meanwhile, the differential cross sections in Figs.~1-3 also depend
on the color-octet matrix elements. As discussed in the last section,
in our calculations, we can only determine the combinations
of the color-octet matrix elements in the form of Eq.~(\ref{combination}).
From the fit, we find the following results for $J/\psi$
and $\Upsilon$ productions:
\begin{eqnarray}
\langle\o^{J/\psi}[{}^1S_0^8]\rangle+\frac{7}{m_c^2}\langle\o^{J/\psi}[{}^3P_0^8]\rangle
&=&0.0197\pm 0.0009   \, {\rm GeV}^3 \ ,  \nonumber \\
\langle\o^{\Upsilon}[{}^1S_0^8]\rangle+\frac{7}{m_b^2}\langle\o^{\Upsilon}[{}^3P_0^8]\rangle
&=&0.0321 \pm 0.0014  \, {\rm GeV}^3 \ , \label{matrix1}
\end{eqnarray}
respectively. It's interesting to note that the result of Eq.~(\ref{matrix1}) agrees well
 with that determined by comparing the total cross section in photoproduction~\cite{Amundson:1996ik}
 and fixed-target hadroproduction~\cite{Beneke:1996tk} of $J/\psi$,
 which are 0.02~${\rm GeV}^3$ and 0.03~${\rm GeV}^3$, respectively.
Comparing these results with those extracted from
the high $P_\perp$ heavy quarkonium production~\cite{Ma:2010yw,Wang:2012is}:
\begin{eqnarray}
\langle\o^{J/\psi}[{}^1S_0^8]\rangle+\frac{3.9}{m_c^2}\langle\o^{J/\psi}[{}^3P_0^8]\rangle
&=&0.074\pm0.019  \, {\rm GeV}^3\ , \nonumber\\
\langle\o^{\Upsilon}[{}^1S_0^8]\rangle+\frac{0.45}{m_b^2}\langle\o^{\Upsilon}[{}^3P_0^8]\rangle
&=&0.113\pm0.020  \, {\rm GeV}^3 \ \,,
\end{eqnarray}
we find that the color-octet matrix element of $[{}^3P_0^8]$ is likely
to take a negative value.
Studying the production of heavy quarkonium in high $P_\perp$ region
 also yields the following relation~\cite{Ma:2010yw,Wang:2012is}:
\begin{eqnarray}
\langle\o^{J/\psi}[{}^3S_1^8]\rangle+\frac{-0.56}{m_c^2}\langle\o^{J/\psi}[{}^3P_0^8]\rangle
&=&0.0005\pm0.00028  \, {\rm GeV}^3\ , \nonumber\\
\langle\o^{\Upsilon}[{}^3S_1^8]\rangle+\frac{-0.045}{m_b^2}\langle\o^{\Upsilon}[{}^3P_0^8]\rangle
&=&0.061 \pm 0.012  \, {\rm GeV}^3 \ . \label{matrix6}
\end{eqnarray}
Combining equations from Eq.~(\ref{matrix1}) to Eq.~(\ref{matrix6}), we deduced:
\begin{eqnarray}
\langle\o^{J/\psi}[{}^1S_0^8]\rangle &=&0.1423\pm0.044  \, {\rm GeV}^3\ , \nonumber\\
   \langle\o^{J/\psi}[{}^3P_0^8]\rangle/m_c^2&=&-0.0175\pm0.0064  \, {\rm GeV}^3\ ,\nonumber\\
   \langle\o^{J/\psi}[{}^3S_1^8]\rangle&=&-0.0093\pm0.0038  \, {\rm GeV}^3 \ ,
   \label{hm1}
\end{eqnarray}
and
\begin{eqnarray}
\langle\o^{\Upsilon}[{}^1S_0^8]\rangle &=&0.119\pm0.021  \, {\rm GeV}^3\ ,  \nonumber\\
   \langle\o^{\Upsilon}[{}^3P_0^8]\rangle/m_b^2&=&-0.012\pm0.0033  \, {\rm GeV}^3\ ,   \nonumber\\
   \langle\o^{\Upsilon}[{}^3S_1^8]\rangle&=&0.060\pm0.012  \, {\rm GeV}^3\ .
   \label{hm2}
\end{eqnarray}
The above result shows that
the color-octet matrix elements can be determined to some degree by
fitting the three linear combinations of the LDMEs to the transverse momentum
distributions of heavy quarkonia
($J/\psi$ and $\Upsilon$) produced in high energy hadron-hadron collisions.

Before closing this section, we note that in the NRQCD
factorization formalism, it has been assumed that the momentum of
the observed quarkonium
is the same as the heavy quark-antiquark pair produced at the short distance,
where the soft gluon radiation from the later stage
({\it i.e.}, hadronization process) was neglected.
In principle, the (non-perturbative) soft gluon radiation
from the final state long distance process
may affect the $P_\perp$ distribution of heavy quarkonium.
 It is worthwhile to further investigate its
effect and implication in heavy quarkonium production in
hadron-hadron collisions.
However, that is beyond the scope of the current paper.

\section{summary and conclusion}

In this paper, we combine the NRQCD and soft gluon resummation formalism to calculate
the $P_\perp$ distribution of $J/\psi$ and $\Upsilon$ production in
hadronic collision in the low $P_\perp$ region. At high energy colliders, the dominant
production mechanism of heavy quarkonium is via gluon-gluon fusion,
similar to the production of Higgs boson at the CERN Large Hadron Collider (LHC).
Our analytic calculation shows that the CSS resummation formalism can be applied to
the $J/\psi$ and $\Upsilon$ hadroproduction processes, similar to the Higgs
 boson hadroproduction.
As compared to the color-singlet Higgs boson hadroproduction,
the coefficient function $B^{(1)}$ of the Sudakov factor, in the CSS resummation formalism,
has to be modified in order to take into account the interference effect
of initial and final state soft gluon radiations, for
the color-octet $J/\psi$ and $\Upsilon$ hadroproductions,
while the coefficients $A^{(1)}$  and $A^{(2)}$ remain to be
the same for the color-singlet and color-octet cases.
In order to numerically evaluate the Sudakov form factor in
the large impact parameter ($b$) region, which is relevant to
the low transverse momentum ($P_\perp$) region, we need to introduce a non-perturbative
function $W^{NP}$ in the CSS resummation formalism. In this work,
we use a 3-parameter pure Gaussian form (BLNY form)
to parameterize the $W^{NP}$, and fit these free parameters by five
experimental data sets of $J/\psi$ and $\Upsilon$ hadroproductions, with a
total number of 26 data points.
We find that we need to modify the non-perturbative
function $W^{NP}$ previously assumed. The result of our analysis is summarized in Table~I.
Though the non-perturbative function $W^{NP}$
extracted from heavy quarkonium data could in principle be applied to studying
the distribution of Higgs boson produced via gluon-gluon fusion process, it should be emphasized that
 the main contribution of the $J/\psi$ and $\Upsilon$ productions in
hadronic collision is from color-octet finial state
channels, in contrast to the production of the color-single Higgs boson.
Hence, they might need different  non-perturbative
factor $W^{NP}$ to describe their low $P_\perp$ distributions.
Nevertheless, it remains useful to compare this newly determined
$W^{NP}$ to that currently used for the Higgs boson and di-photon
productions in hadronic collisions.

As emphasized in the above discussions, our resummation
calculation for the heavy quarkonium production was
derived from a NLO result in the RQCD framework.
Because of the simple color configuration and non-relativistic
nature in this formalism, the soft gluon radiation can be resummed
into a simple exponential form. This feature is consistent with the
soft gluon radiation in heavy quark pair production previously studied
in the partonic threshold limit with heavy quark pair produced
at rest. In particular,
the matrix form of the Sudakov form factor can be simplified in
this limit, which is consistent to our results. This is encouraging
for further investigations on the transverse momentum resummation
for more complicated hard processes. Of course, we have to keep
in mind that the resummation formula may break down at higher orders
because there has been not general proof of the factorization for
hardronic hard processes (such as dijet, heavy quark pair,
and heavy quarkonium productions) in hadron collisions.

We have also shown how to extract the values of the
 color-octet matrix elements from studying the
 transverse momentum distribution of $J/\psi$ and $\Upsilon$, in
 both the low and high $P_\perp$ regions. The result of this analysis is
 given in Eqs.~(\ref{hm1}) and (\ref{hm2}).
These matrix elements are found to be consistent with the inclusive
total cross section in photoproduction~\cite{Amundson:1996ik}
and fixed-target hadroproduction~\cite{Beneke:1996tk} of $J/\psi$. Further investigations are
needed to clarify the underlying mechanisms for heavy quarkonium
production in the whole $P_\perp$ range.

We would like to emphasize that the resummation formula in
our calculations are based on the extension of the NRQCD factorization
in the low transverse momentum region, where the most  singular contributions
are found to follow the CSS resummation expansion at one-loop order.
It will be interesting to check if this holds at higher orders.
Meanwhile, we note that the color-single ${}^3S_1^{(1)}$ channel
does not contain singular contributions at low transverse momentum,
because it cannot be produced via a $2\to 1$ process. Hence, we
did not consider the effect of soft gluon resummation for this production
channel in the current analysis.

\section{acknowledgements }

We thank J.~.P.~Ma, J.~W.~Qiu, and in particular, K.~T.~Chao and his group at Peking University
for the discussions and comments. We also thank J.~C.~Peng for discussions on the E866 experimental
results. This work was partially supported by the U.
S. Department of Energy via grant DE-AC02-05CH11231, and by the U.S. National Science Foundation
under Grant No. PHY-0855561.


\end{document}